\documentclass[a4paper,11pt]{article}
\usepackage{jinstpub} 

\usepackage{color}
\usepackage[utf8]{inputenc}

{\title{\boldmath Revisiting the ARM cut in Compton gamma-ray imaging and its application to the INSPIRE detector}}

\author{J.Kataoka$^a$, S.Ogasawara$^a$, R.Mori$^a$, 
K.Yamamoto$^a$, A.R.Joshi$^a$, S.Kojima$^a$, K.Sato$^a$,
 K.S.Tanaka$^a$, K.Watanabe$^b$, M.Yasuda$^b$, H.Kobayashi$^b$,
D.Kobayashi$^b$, A.Ohira$^b$, Y.Amaki$^b$, Y.Arai$^b$, K.Tashirio$^b$, K.Otsubo$^b$, Y.Ozeki$^b$, Y.Kawaguchi$^b$, D.Yoshimura$^b$, H.Yoshida$^b$, K.Takahashi$^b$, S.Masaki$^b$, 
N.Yamada$^b$, K.Oikawa$^b$, E.Zamami$^b$, 
K.Miyamoto$^b$, T.Chujo$^b$, H.Nakanishi$^b$, T.Tomura$^b$, S.Hayatsu$^c$, M.Fukuda$^c$, H.Seki$^c$, S.Joshima$^c$, Y.Yatsu$^c$}
\affiliation{$^a$ Faculty of Science and Engineering, Waseda University, 
Tokyo, 169-8555, Japan}
\affiliation{$^b$ School of Engineering, Institute of Science Tokyo, Tokyo, 152-8550, Japan}
\affiliation{$^c$ School of Science, Institute of Science Tokyo, Tokyo, 152-8550, Japan}

\emailAdd{kataoka.jun@waseda.jp, shojun02-oubutsu@suou.waseda.jp}

\abstract{The Compton camera is a gamma-ray imaging device developed in the 1970s. In the 1990s, the COMPTEL detector onboard the CGRO was the first to utilize a Compton camera for MeV all-sky survey observations. Recently, various Compton cameras have been developed using scintillators, semiconductors, and gas detectors, some of which are intended for future small satellite missions as well as medical applications.  However, the image obtained by a Compton camera has strong artifacts owing to the overlap of the Compton cones or the arcs, which degrade the resolution and sensitivity of the image. In this study, we revisit the adaptive ARM cut that significantly reduces artifacts when the direction of gamma ray emitting source is already known. This approach complements the statistically well-defined method based on the response function in the three-dimensional data space of scattering direction ($\chi$, $\psi$) and scattering angle $\theta$, but it is more direct, intuitive, and simplifies the extraction of spectra in astronomical observations of point-like sources.
Using a Compton camera,  INSPIRE,  onboard the ultra-small satellite GRAPHIUM as an example, 
we numerically evaluated the extent of 
background reduction to   estimate simulation-based sensitivity. The method was also applied to actual measurements using a quarter-scale prototype of INSPIRE to  extract spectra from multiple sources within the same field of view. 
}


\begin{document}
\maketitle
\flushbottom

\section{Introduction}
Gamma rays with energies of a few hundreds of keV to several MeV are important probes for exploring nucleosynthesis in the universe, heating of interstellar matter, chemical evolution, and cosmic ray acceleration. However, gamma rays above MeV cannot be focused by lenses or mirrors and are difficult to image with coded masks \cite{and23}. The Compton camera, which uses Compton kinematics, was invented in the 1970s aiming for both the astrophysical and medical  applications \cite{sh73,to74}. Especially 
in the 1990s, the CGRO satellite COMPTEL detector \cite{sh93} conducted the first all-sky survey observations in the 0.75$-$30 MeV range. COMPTEL detected 31 gamma-ray sources, including active galactic nuclei (AGN) and pulsars,  and diffuse gamma rays along the Galactic plane originating from Al-26 \cite{sh00}. However, the sensitivity achieved by COMPTEL was approximately two orders of magnitude lower than those for other wavelengths. Thus, MeV gamma-ray astronomy is still in its infancy.

In a Compton camera, gamma rays scattered by a scatterer are captured by an absorber, and the Compton scattering angle $\theta$ is calculated from the energy deposited to each. When the electron recoil direction is unknown, incident direction of the gamma ray is restricted to a cone surface with a half apex angle $\theta$ with the line connecting the reaction positions of the scatterer and absorber, whose direction is defined by two orthogonal coordinates ($\chi$,$\psi$). 
The actual direction of gamma ray source 
can be determined by superposing such ``Compton cones"  in back-projection image. Meanwhile, the overlap of the cones produces a strong artifact like skirts in the back-projection image, even when observing a point-like source. Moreover, if a scattered gamma ray is not fully absorbed, the resulting Compton cone may indicate an incorrect direction, further degraging image quality.
However,  in most medical imaging 
applications, a simple superposition of Compton cones is sufficient,  as  escape events can be effectively eliminated using an  ``energy cut" when   target radioisotopes, i.e., energy of gamma-ray lines, are known. 
In contrast, such simple approach is generally not applicable  in the analysis 
of COMPTEL data,  where  
background typically dominates the source signal and the observed spectra are almost  featureless \cite{sh93}. 
COMPTEL comprises large liquid and NaI(Tl) scintillators; recently, various Compton cameras have been proposed, including semiconductors like CdTe and HPGe \cite{wa14,to24} for applications to medical imaging and environmental survey. In particular, some gas detectors and/or Si pixel detectors \cite{ta04,ta22,yo18} may track the direction of recoil electrons, thereby restricting a part of the Compton cone as an arc, i.e., a certain fraction within the cone. However, the detection efficiency of electron-tracking Compton cameras is typically lower than that of  scintillator-based detectors, and artifacts remain even for such arcs overlaps.

The Angular Resolution Measure (ARM) is typically used to evaluate the image resolution of  Compton cameras. The ARM corresponds to the deviation between $\theta_{\rm src}$ and $\theta_{\rm E}$, that is, ARM = $\theta_{\rm src}$ $-$ $\theta_{\rm E}$, where 
$\theta_E$ is the scattering angle calculated from the measured energy deposit, and 
$\theta_{\rm src}$ is calculated from the measured interaction position and true direction of the source. The three major factors that determine ARM  are the energy resolution of the detector, 
the accuracy of determining the reaction position, and Doppler broadening\cite{zo03}, which corresponds to the momentum fluctuation of the scattered electrons. For example, in commercially available Compton cameras, an ARM distribution of 5$-$10$^{\circ}$ (FWHM) is typical for 662 keV gamma rays (e.g., \cite{kat13,kat18}). However, the actual image exhibits more broadening than  what expected from such an ARM distribution because of aforementioned artifacts.
Yet, a constraint based on ARM distribution is essential for verifying the consistency of Compton kinematics and thus for eliminating  background events. In fact, such a event  selection is implemented in a convolutional manner within the COMPTEL response function\cite{sh93}, but the COMPTEL Processing and Analysis Software (COMPASS)\cite{de94} was only available at the COMPTEL collaboration institutes and was specific to analyze the COMPTEL data.

This study reviews a simple and practical method for applying an ARM cut to effectively  reduce background events in future astronomical observations.  Similar to  conventional approaches in medical imaging, Compton cones are simply superposed and projected on the sky,  but with the additional application of an 
ARM constraint. 
The  reviewed method is evaluated for its ability to enhance sensitivity in pointed observations of known sources such as AGN and pulsars. In particular, we demonstrate that the method is effective for extracting the spectra of multiple sources within the same FOV, provided that the separation between sources is sufficiently larger than the typical scale of the ARM distribution. 
We are developing an ultra-small satellite GRAPHIUM, which is to be launched in 2027. 
GRAPHIUM is equipped with INSPIRE \cite{kat24}, a wideband X-ray and gamma-ray camera, and 
its performance is being evaluated using an engineering model (EM). In this paper, we numerically and experimentally evaluated the extent to which the sensitivity and resolution could be improved by applying the reviewed method to INSPIRE.

\section{Gamma-ray imaging by a Compton camera}
\subsection{Event selection by energy cut}
In a Compton camera, if an incident gamma ray of energy $E$ deposits a part of energy ($E_1$) at the position $X_1$ in the scatterer and the remaining energy ($E_2$) in the absorber at the position $X_2$, the scattering angle $\theta_E$, determined from the energy deposited can be expressed 
as follows:

\begin{equation}
cos \theta_E = 1 - \frac{m_e c^2}{E_2} + \frac{m_e c^2}{E_1 + E_2}, 
\end{equation}
where $m_e c^2$ is the rest mass energy of electrons.

Thus, the direction of arrival gamma rays is restricted to the surface of a Compton cone with 
a half apex angle $\theta_E$ with respective to the straight line connecting $X_1$ and $X_2$. 
The most likely position or direction of the radiation source can be identified by overlapping 
the cones for several events. This is the basis of gamma-ray imaging using a Compton camera.
In particular, when the energy of such a source is known like nuclear medicine imaging, the signal-to-background (S/B) ratio of the image can be improved by applying an energy cut either to $E_1$ or $E_2$ and to their sum, $E$. 

\begin{figure}[ht]
    \centering
    \includegraphics[width = 16.0cm]{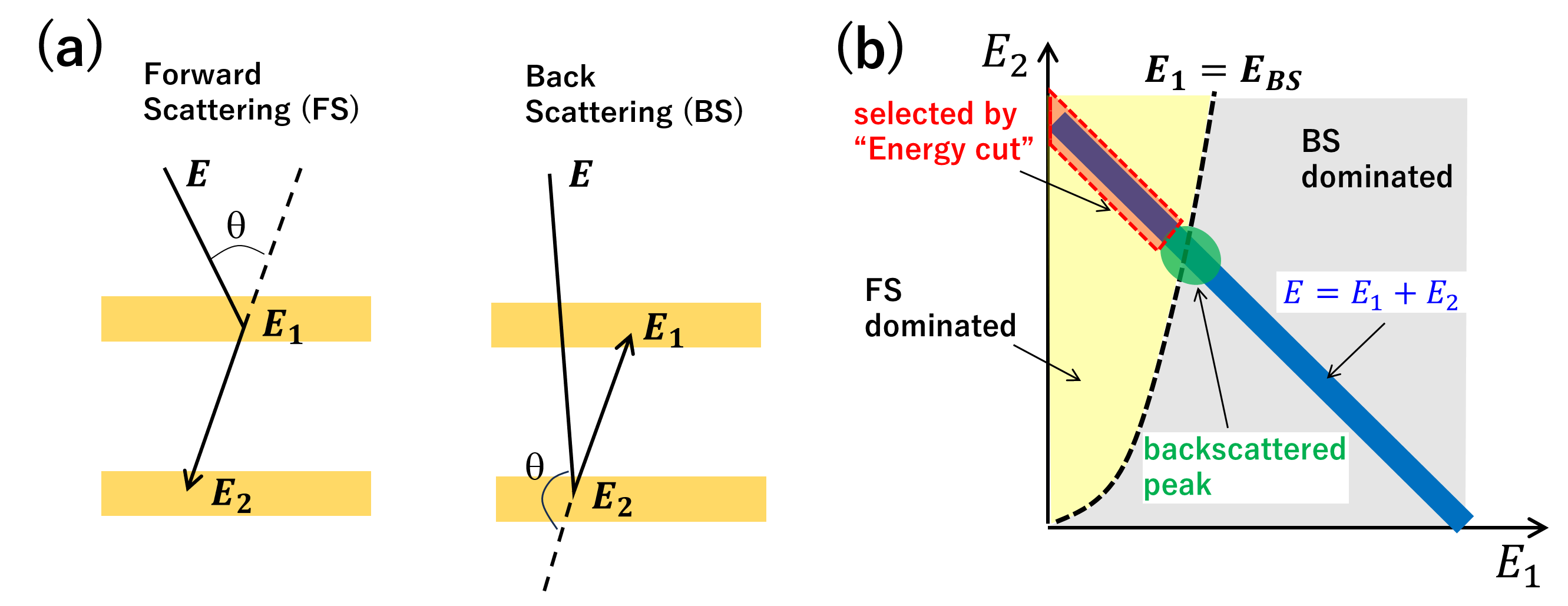}
    \caption{(a) Schematic of Compton scattering in the forward and backward directions. (b) Two dimensional diagram of $E_1$ and $E_2$ and the typical region  for energy cut imaging.}
    \label{fig:Ecut}
\end{figure}

In many nuclear medicine imaging applications, e.g.,\cite{ki17}, images are  generated by selecting forward scattering (FS) events (Fig.\ref{fig:Ecut}(a)).
Because the energy deposit corresponding to the backscattered  (BS) peak is given by  
\begin{equation}
E_{\rm BS} = \frac{E}{1+\frac{2E}{m_e c^2}}, 
\end{equation}
we can effectively eliminate backscattered events by applying energy cut $E_1$ $\le$ $E_{\rm BS}$, 
where $E_1$ represents the energy deposit in the first layer and $E_2$ in the second layer. For example, energy cut like $E_1$ 
$\le$ 184 $-$ $\Delta{E_1}$~keV and 
662$-$$\Delta E$~keV $\le$ $E$ $\le$ 662+$\Delta E$~keV are often applied while imaging 
a Cs-137 source which emits 662 keV gamma rays, where $\Delta$$E$ and $\Delta{E_1}$ 
correspond to the energy resolution of detector (Fig.\ref{fig:Ecut}(b)). 
But note that, the above selection is correct only for a two-hit and complete Compton sequence. Some FS events are excluded when a part of the gamma-ray energy are not deposited in the detector (i.e., escape events). Conversely, some BS events may be wrongly interpreted and selected as FS events. Thus we regard this event selection just tentative, as referred to "FS dominated" and "BS dominated "
regions in Fig.\ref{fig:Ecut}(b).

Even when multiple radiation sources emitting gamma rays of different energies exist 
in the same FOV, an image of each source can be extracted by adopting multiple energy cuts (e.g., \cite{ki17}). 
However, extracting the spectrum by selecting a certain region of the image where the source may exist is almost impossible because the overlap of Compton cones severely contaminates the image formed from different radiation sources.

Furthermore, in the case of astronomical gamma-ray observations, no characteristic line emission occurs, except for nuclear gamma rays, such as Al-26 and 511 keV annihilation gamma rays as mentioned above.
For example, most AGNs, pulsars, and gamma-ray bursts exhibit featureless non-thermal spectra that are typically represented by a  power-law with a spectral index ($\Gamma$ of $E^{-\Gamma}$) between 1.0 to 3.0. Even in such  continuum emissions, 
we can adopt an appropriate energy-cut for each event by comparing $E_{BS}$ and $E_1$,  as described above.  For example, backscattered events are effectively removed 
in the analysis of gamma-ray imaging of thunderclouds, 
whose gamma-ray spectrum follows a power-law function of  $\propto$ $E^{-1.6}$ \cite{ku22}.

\subsection{{Adaptive ARM cut}}
\begin{figure}[ht]
    \centering
    \includegraphics[width = 10.0cm]{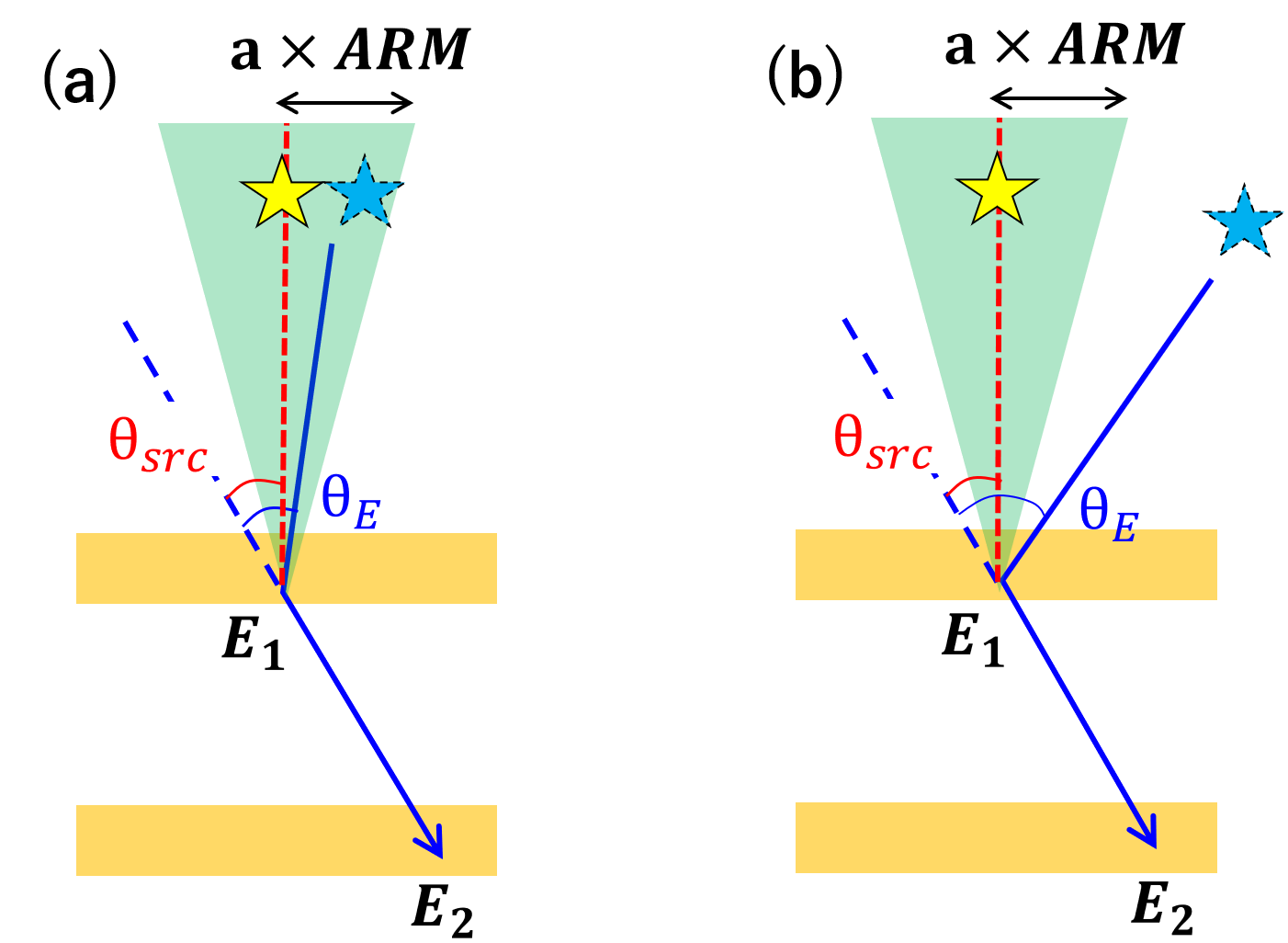}
    \caption{Schematic of ARM cut.  Only the events that  most probably originated from the pointing source can be distinguished.  (a) source-like events (b) background events.}
    \label{fig:Acut}
\end{figure}

In medical imaging,  the aforementioned energy cut is 
effective for extracting the  distribution of radioisotopes, that emit monochromatic line gamma rays.  In contrast, many astronomical observations  involve  pointing at well-known sources, such as  AGN and pulsars, where the spatial images are  trivial,  but extracting high-quality spectral is critically important. 
The concept of adaptive ARM cut is illustrated in Fig.\ref{fig:Acut}. When the $\theta_E$ calculated from Eqn.(2.1) matches the true source direction ($\theta_{\rm src}$) within a certain range defined by the ARM distribution, the event is most likely to have originated from the celestial source. Thus, 

\begin{equation}
 \theta_{\rm src} - a\times{{\rm ARM_{FWHM}}(E)}\le \theta_E \le \theta_{\rm src} + a\times{{\rm ARM_{FWHM}}(E)},
 \end{equation}
where ${\rm ARM_{FWHM}}(E)$ is the energy-dependent FWHM value of ARM distribution and $a$ is a constant,  typically $a$ $\simeq$ 1. Events that do not meet the aforementioned condition are regarded as background events. 

Such event selection is a powerful tool for  significantly reducing the contamination of isotropic backgrounds, particularly for  wide-field imagers such as Compton cameras.   
In principle, an energy cut is not required, as described in $\S$2.1; thus,  the number of events that can be used for image reconstruction increases, as even forward-scattered events in the BS-dominated region can also be used for analysis. Furthermore, even when multiple sources are present within the same field of view (FOV), the spectrum of each source can be extracted with minimal contamination, provided that the separation angle exceeds the scale of the ARM distribution. However, this simple analysis may not be applicable for faint objects when bright sources are located nearby within the same FOV. In such cases, statistical data analysis, as demonstrated in \cite{kn22}, becomes essential.

Note, that, in the case of COMPTEL data analysis, a three-dimensional space consisting of the coordinates ($\chi$,$\psi$) of the celestial sphere and scattering angle $\theta$ is defined, and a segmented response function is prepared. In the ideal case where scattered gamma rays are fully absorbed in the absorber,  the gamma-ray 
pattern from a  celestial coordinate  ($\chi_0$, $\psi_0$) lies on a surface in the  
($\chi$,$\psi$, $\theta_E$) space, where the apex of  the 
cone  is located at ($\chi_0$, $\psi_0$)  and the cone's  semi-angle is 45$^{\circ}$. However,  incompletely absorbed events tend to populate the interior of the cone.
To reconstruct the spatial distribution that best explains the observed data, methods such as Maximum Entropy  \cite{st92} and Maximum Likelihood methods \cite{bo92} have been used, producing images in various energy bands $-$ typically in 0.75$-$1, 1$-$3, 3$-$10 and 10$-$30~MeV.   
Whether a photon  scattered in   direction ($\chi$, $\psi$) with scattering angle $\theta_E$ is consistent with a given celestial  position  ($\chi_0$, $\psi_0$)  can be evaluated  using the ARM.   The S/B ratio is further improved by identifying the gamma-ray interaction sequence $-$ either FS or BS events, $-$ using Time-of-Flight (ToF)  information,  and by suppressing neutron background through the pulse-shape discrimination.

\section{Application to INSPIRE onboard GRAPHIUM}
\subsection{Overview of the detector}
\begin{figure}[ht]
    \centering
    \includegraphics[width = 14.0cm]{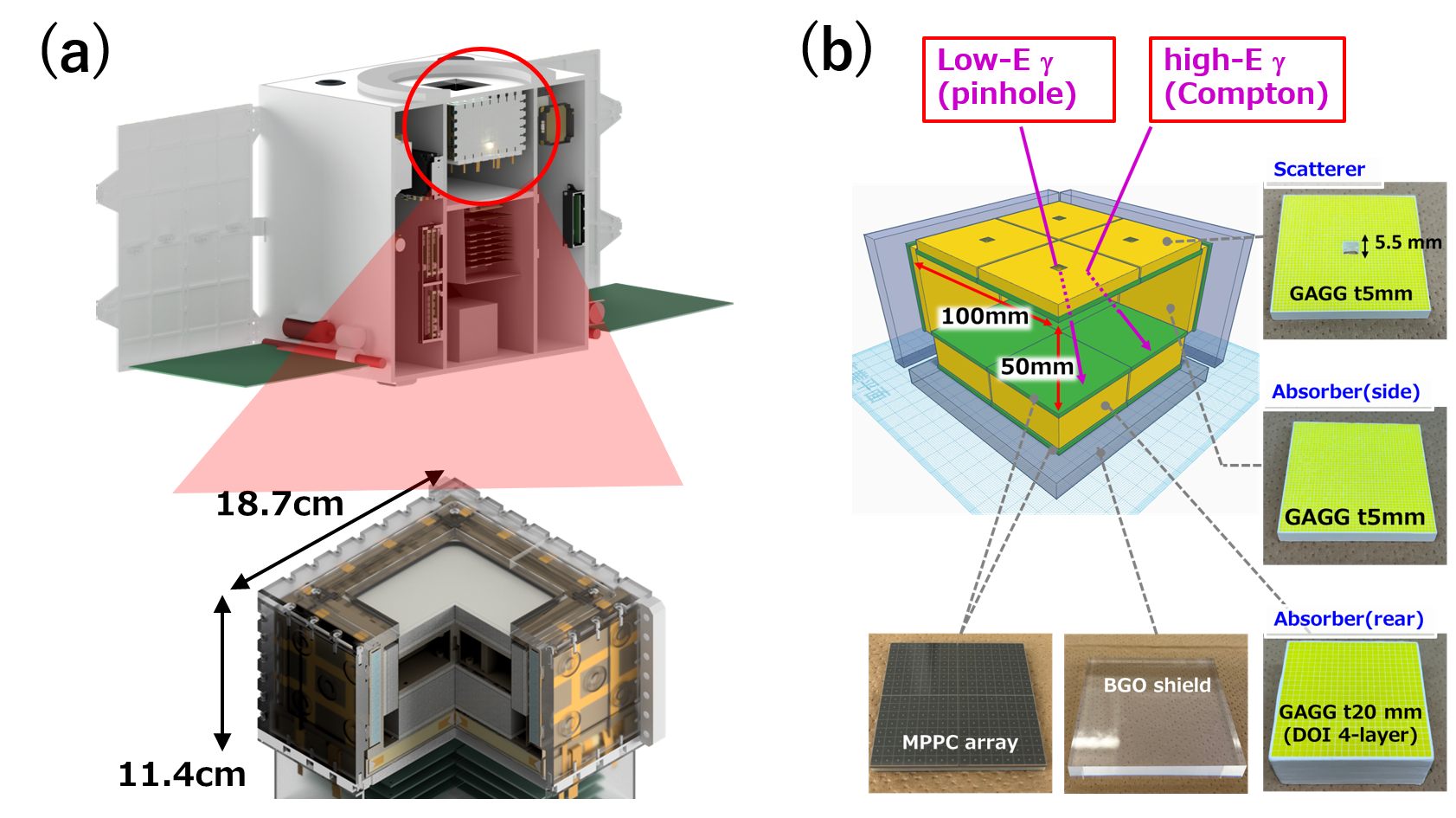}
    \caption{(a) Overview of the GRAPHIUM satellite and the wideband X-ray and gamma-ray camera INSPIRE. (b) Concept of HCC.}
    \label{fig:INSPIRE}
\end{figure}

Waseda University, in collaboration with Science Tokyo, is developing an ultra-small satellite, GRAPHIUM (Fig.\ref{fig:INSPIRE}(a)), which is expected to be launched in 2027. The dimensions of GRAHIUM are 60$\times$60$\times$60~cm$^3$, and its approximate weight is 65~kg. The assumed  orbit is the Sun Synchronous Orbit (SSO) at 10$-$12~hr LST. GRAPHIUM is equipped with a wide-field and broadband X-ray and gamma-ray camera INSPIRE for all-sky monitoring between 30~keV and 3~MeV to obtain 
new information for MeV gamma-ray astronomy. INSPIRE achieves a detector area of 10$\times$10~cm$^2$ 
via four hybrid Compton cameras (HCC) of 5$\times$5~cm$^2$ \cite{omt20}. Its 
sensitivity is further improved by placing absorbers on the sides. Total weight is approximately 10~kg. The overall dimensions of INSPIRE are shown in Fig.\ref{fig:INSPIRE}(b).

\begin{figure}[ht]
    \centering
    \includegraphics[width = 15.3cm]{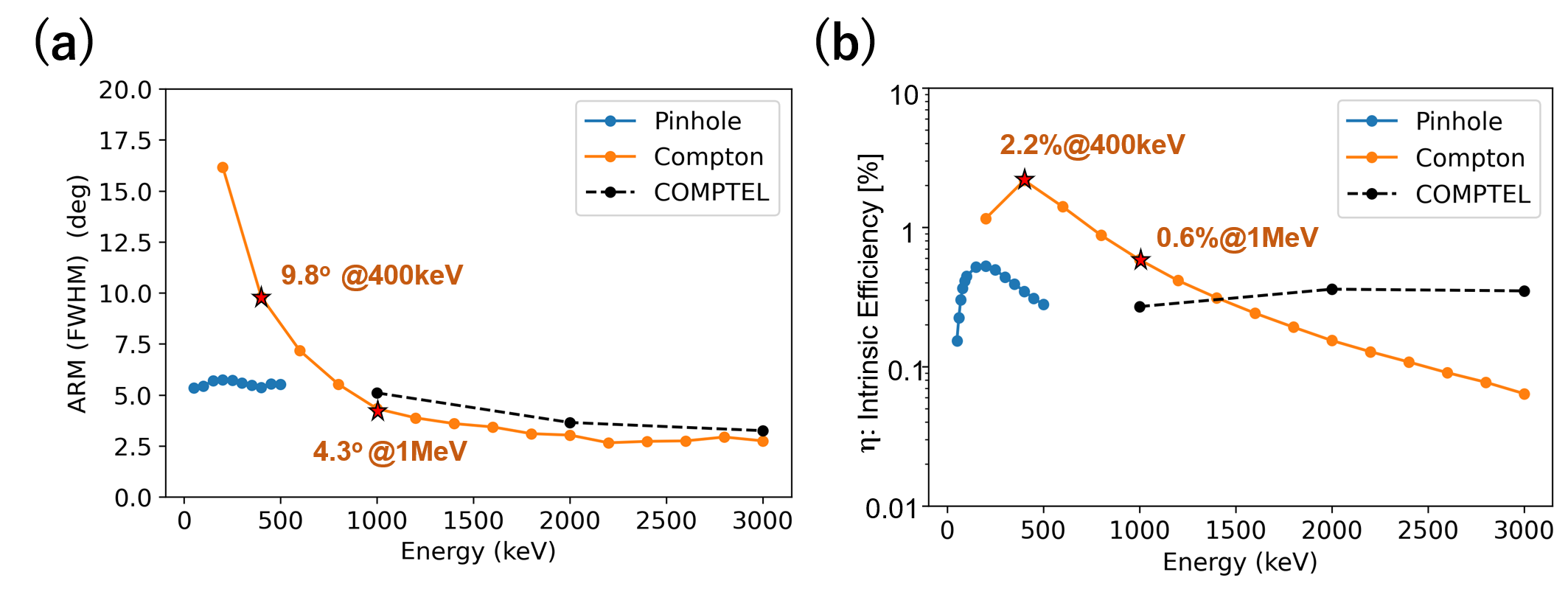}
    \caption{(a) Angular resolution for the pinhole mode, ARM for Compton mode, and (b) intrinsic efficiency of INSPIRE.}
    \label{fig:angeta}
\end{figure}

The HCC is composed of a scatterer and an absorber that combines a pixelized GAGG scintillator array coupled with an MPPC array and has a small pinhole in the center of the scatterer. Thus, X-rays and low-energy gamma rays of 30$-$200~keV can be imaged as a pinhole camera, whereas gamma rays of 200~keV or more can be imaged as the Compton camera 
(Fig.\ref{fig:INSPIRE}(b)). While all multiple-interaction events are recorded, only two-hit events are used in the Compton data analysis.
The observation FOV is quite large, approximately 1 str  and more than 3 str (or 1/4 of the entire sky)  in the pinhole and Compton modes, respectively,  making it suitable for observing transient objects, such as gamma-ray bursts (GRBs). In addition, the sides and bottom of the detector are covered with BGO scintillator plates to reduce the background and efficiently remove escape events that are not completely absorbed by the absorber. 
The expected angular resolution (FWHM of ARM distribution for the Compton mode) and intrinsic efficiency $\eta$ calculated for the conventional energy cut (see, $\S$2.1) are summarized in Fig.\ref{fig:angeta}. Here, $\eta$ is defined as the fraction of photons entering the scatterer that undergo Compton scattering and are subsequently absorbed in the absorber, enabling event reconstruction. 

\subsection{Crab Nebula simulation with orbital background}
In the case of Low Earth Orbit (LEO), such as SSO, various backgrounds affect the observations on the orbit, which considerably degrades the sensitivity in the MeV range.  In addition, the activation of detectors owing to charged particles (mostly protons) trapped in the orbit is a problem. In particular, in the energy band targeted by INSPIRE, the cosmic X-ray background (CXB) is dominant 
below 150 keV, whereas the albedo gamma rays generated by the interaction of cosmic rays with the atmosphere are dominant as isotropic incident backgrounds above 150 keV. The CXB spectrum is well approximated by a power-law with an index $\Gamma$ $\simeq$ 2.9 above 30 keV \cite{tu10}, and the albedo is approximated by a power-law of $\Gamma$ $\simeq$ 1.3 in the MeV energy band \cite{miz04}. These backgrounds are typically several orders of magnitude higher than the fluxes of any astronomical sources. Thus, achieving a narrow FOV with either passive or active collimator can effectively reduce  the background by several hundred keV. 
This is a common concept employed  in OSSE/CGRO\cite{cam92}, Suzaku/HXD\cite{fuk09},  and more recently, in the soft gamma-ray detector (SGD)\cite{taj10} onboard the Hitomi satellite.
However, above the MeV range, blocking such a background is difficult because the necessary shield becomes too thick and heavy. Background rejection posed the biggest challenge for COMPTEL with an FOV of 1 str, and such situation worsened in the case of INSPIRE with a wider FOV of 3 str.

\begin{figure}[ht]
    \centering
    \includegraphics[width = 10.0cm]{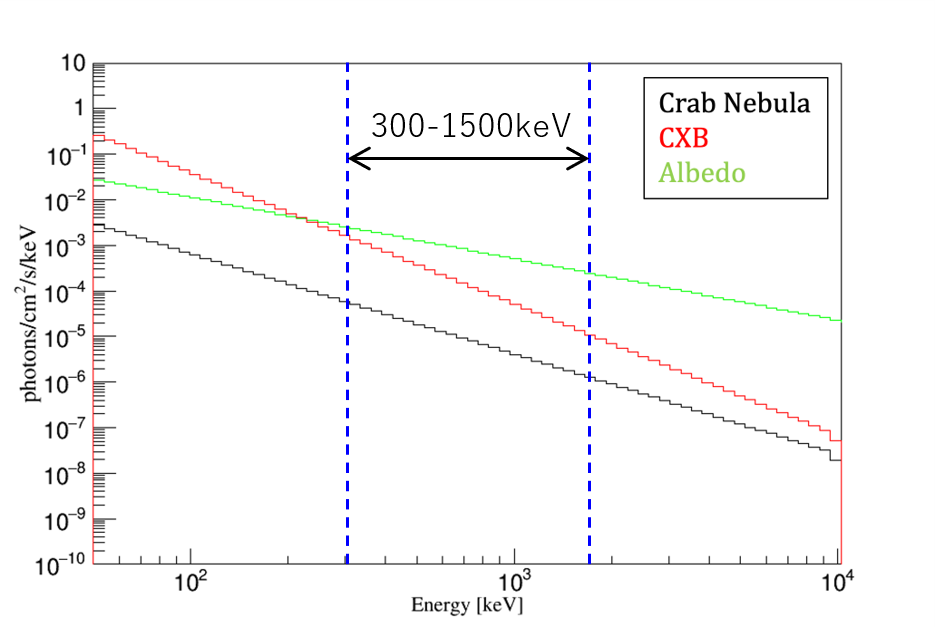}
    \caption{Comparison of X-ray and gamma-ray spectra of 
    the Crab Nebula ($black$), CXB ($red$) and albedo gamma rays ($green$). The fluxes of CXB and albedo gamma rays are integrated over all solid angles.} 
    \label{fig:BGD}
\end{figure}

Here, we used Geant~4 \cite{ago03} to first perform an observational simulation of the Crab Nebula, assuming an isotropic background.
The angular extent of the Crab Nebula is 0.2$^{\circ}$; therefore, it can be regarded as a point-like source compared to the angular resolution of INSPIRE, whose ARM distribution is 4.3$^{\circ}$~(FWHM) at 1~MeV (Fig.\ref{fig:angeta}(a)). In the simulation, the Crab Nebula, whose energy spectra follows a  power-law with an index $\Gamma$ of 2.1 \cite{jo09}, was pointed at the center of FOV, and the CXB and albedo gamma rays were randomly shot to mimic their actual intensities and spectra \cite{miz04}. The energy and position uncertainties of the actual detector are all taken into account when deriving the image and spectrum. The assumed incident spectra for each component are shown in Fig.\ref{fig:BGD}, where the CXB and albedo gamma ray show the flux integrated over the total solid angle (i.e., 4$\pi$ str). Although the Crab Nebula is one of the brightest sources in the sky, the fluxes of the CXB and albedo are always several orders of magnitude higher in the MeV range. In addition, the activation of the GAGG and BGO scintillator in orbit will provide additional background, which was neglected in the following simulations.  In fact, proton irradiation experiments, assuming primary protons and trapped protons during the South Atlantic Anomaly (SAA) passages,  indicated that the background contribution due to activation was less than 30–50 $\%$ of the CXB plus albedo gamma rays across the entire energy band; however, the flux of the 511 keV gamma-ray increased by 60 $\%$   (Yamamoto et al. {\it submitted}). Therefore, the activation of INSPIRE did not significantly affect subsequent simulations. We will revisit this issue in $\S$4.2.  
In addition, we limit our discussion in the Compton mode to the application of the  ARM cut for imaging and extracting the spectrum.

\subsection{Energy cut vs ARM cut: imaging} 
\begin{figure}[ht]
    \centering
    \includegraphics[width = 14cm]{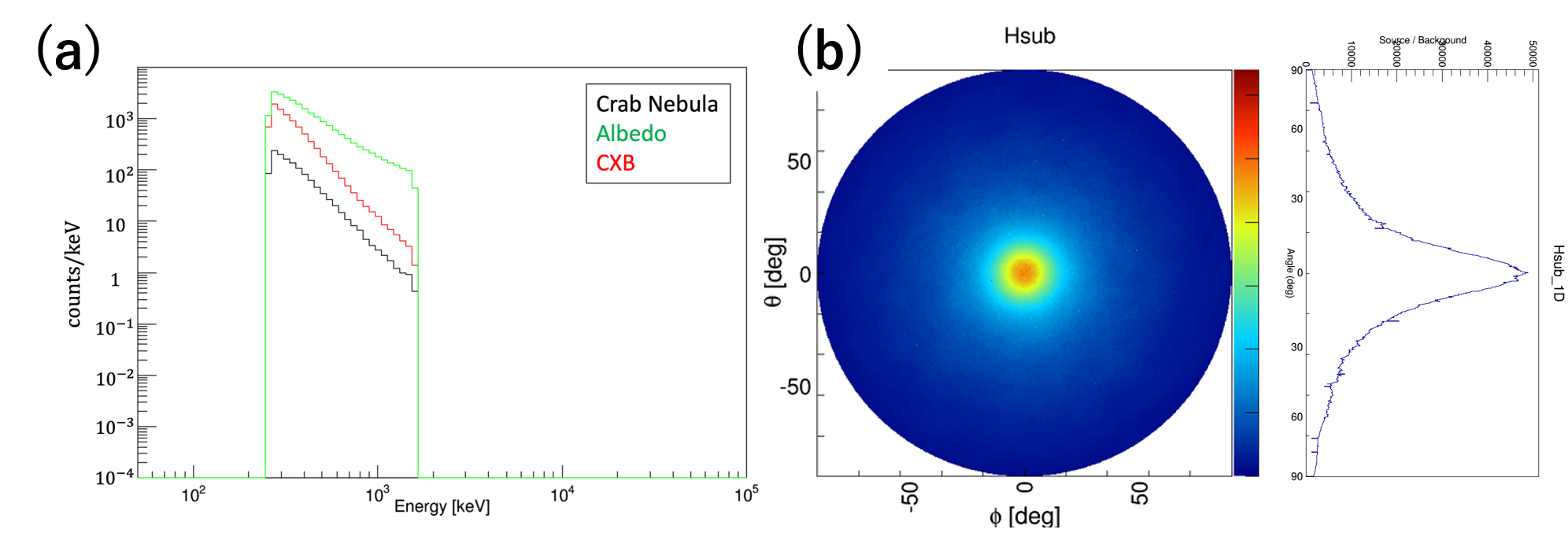}
    \caption{Simulation of the 1~Msec INSPIRE observation of the Crab Nebula under the energy cut method. (a) Comparison of the spectra after energy cut 
    (b) back-projection image after subtracting the OFF-source image from ON-source image.} 
    \label{fig:EcutCrab}
\end{figure}

Here, we assumed a 1~Msec (10$^6$ sec) INSPIRE observation of the Crab Nebula to compare images extracted using the energy cut and ARM cut methods. We imaged in the energy band from 300~keV to 1.5~MeV, as indicated by the arrow in  Fig.\ref{fig:BGD}.
First, we applied energy cut, as shown in  Fig.\ref{fig:Ecut}. 
Only forward scattered events between scatterer and absorber 
($E_1$ $\cap$ $E_2$ ) were used, which satisfied $E_1$ $\le$ $E_{\rm BS}$ and did not trigger any BGO signals. 
Note that the energy cut selection reduced background contamination by approximately a factor of five by eliminating events originating outside the FOV; however, it still remained about an order of magnitude higher than that of the Crab Nebula. The expected back-projection image and spectra pointed toward the Crab Nebula is shown in  Fig.\ref{fig:EcutCrab}.
To enhance the source contribution, we subtracted the OFF-source image from the ON-source image, where the Crab Nebula was pointed at the center of the FOV. The right panel is the projection of the image along the vertical axis. Owing to the 
contamination of Compton cones from background events, the FWHM of the Crab enhancement was 29.7$^{\circ}$, which was 
much broader than that expected from the ARM distribution.  This is due to both background cone contamination and inherent cone artifacts, as described in $\S$1.

\begin{figure}[ht]
    \centering
    \includegraphics[width = 14.0cm]{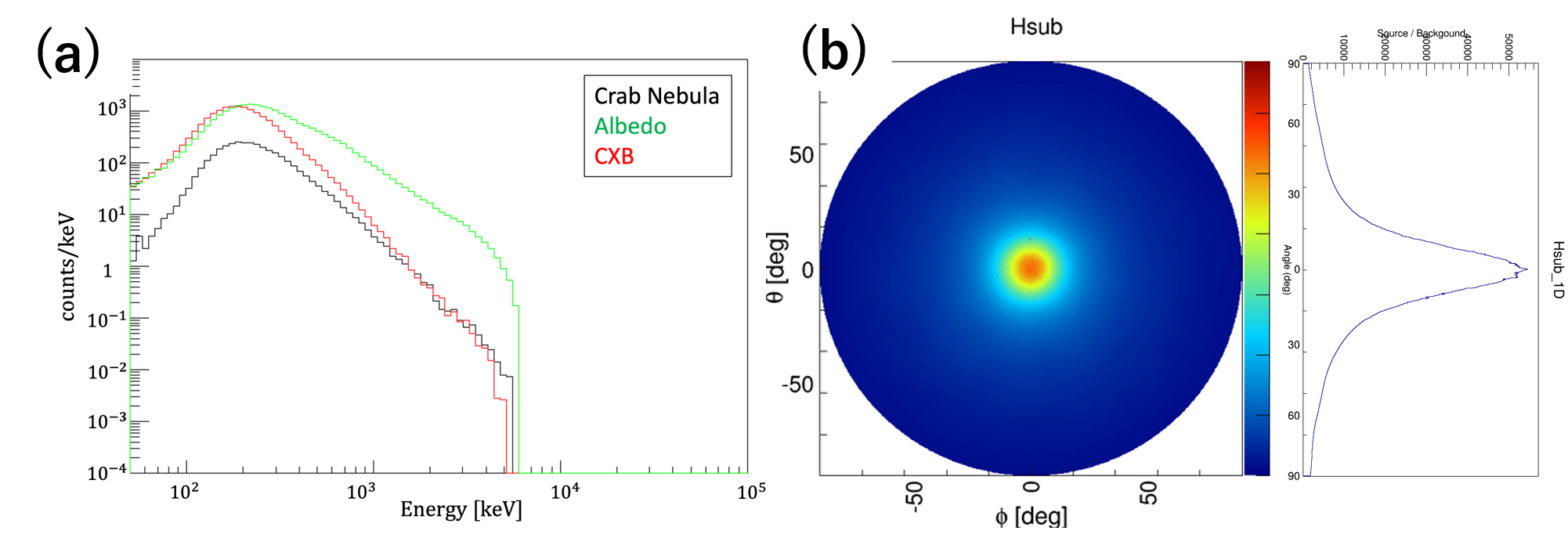}
\caption{Simulation of 1~Msec INSPIRE observation of the Crab Nebula under ARM cut method. (a) Comparison of the spectra after ARM cut    (b) back-projection image after subtracting the OFF-source image from ON-source image.} 
    \label{fig:AcutCrab}
\end{figure}

Next, we performed the same simulation as that applied to the ARM cut.  We applied an ARM cut following 
the energy dependence shown in Fig.\ref{fig:angeta}(a), for example,   9.8$^{\circ}$~(FWHM) at 400~keV and 4.3$^{\circ}$~(FWHM) at 
1~MeV. Additionally, we set $a$ = 1. 
Fig.\ref{fig:AcutCrab}(a) compares the Crab Nebula,  background 
CXB,  and albedo gamma-ray spectra. The S/B ratio of the spectrum is significantly improved compared to that in Fig.\ref{fig:EcutCrab}.  Moreover, the number of 
events selected by the ARM cut is 1.8 times larger than the energy cut. 
Fig.\ref{fig:AcutCrab}(b) shows the ON$-$OFF source image  
as described above, just for comparison with that obtained with the energy cut. All coincidence events ($E_1$ $\cap$ $E_2$ ) were used in the energy range between 300~keV and 1.5~MeV. The Crab Nebula appears more prominently in the image, with an FWHM of 26.7$^{\circ}$.  This is because the ARM cut suppresses the contamination 
from the CXB and albedo gamma rays,  which  
contribute to the tail (or offset) in the off-center image. Note, however,  that  both the 
ON and OFF-source image are significantly  biased toward the center of the image (i.e., the direction of the point source) due to the  nature of the ARM cut  selection. Therefore, the ARM cut proposed in this section  is useful primarily for spectral extraction. 

\subsection{Energy cut vs ARM cut: sensitivity} 
We demonstrate that an ARM cut of $a$ = 1 effectively improves detection significance of the Crab Nebula with increased event statistics, which indicates reduction of  the background and effective improvement of the sensitivity of the detector in the pointing observations.

The detection sensitivity is expressed as a function of the incident photon energy $E$  by the following formula for both the continuum ($S_{\rm C}$) and line components ($S_{\rm L}$):

\begin{equation}
S_{\rm C}(E) = \frac{f}{\eta (E)}\sqrt{\frac{b(E)}{A \Delta E T}}, 
\end{equation}
\begin{equation}
S_{\rm L}(E) = \frac{f}{\eta (E)}\sqrt{\frac{2b(E)\delta E}{AT}}, 
\end{equation}
where $\eta (E)$ is an intrinsic efficiency; $b(E)$ is the background including both the CXB and albedo gamma rays; A = 100~cm$^2$ is the geometrical area of INSPIRE; and $T$ is the observation time in seconds. $\Delta E$ is the energy window for continuum emission, which was set as $E$ = $\Delta E$, and $\delta E$  = 57$\times$($E$[MeV])$^{1/2}$ keV is the energy resolution for line emission. $f$ is the detection significance and we temporarily set $f$ = 3. 

\begin{figure}[ht]
    \centering
    \includegraphics[width = 15.3cm]{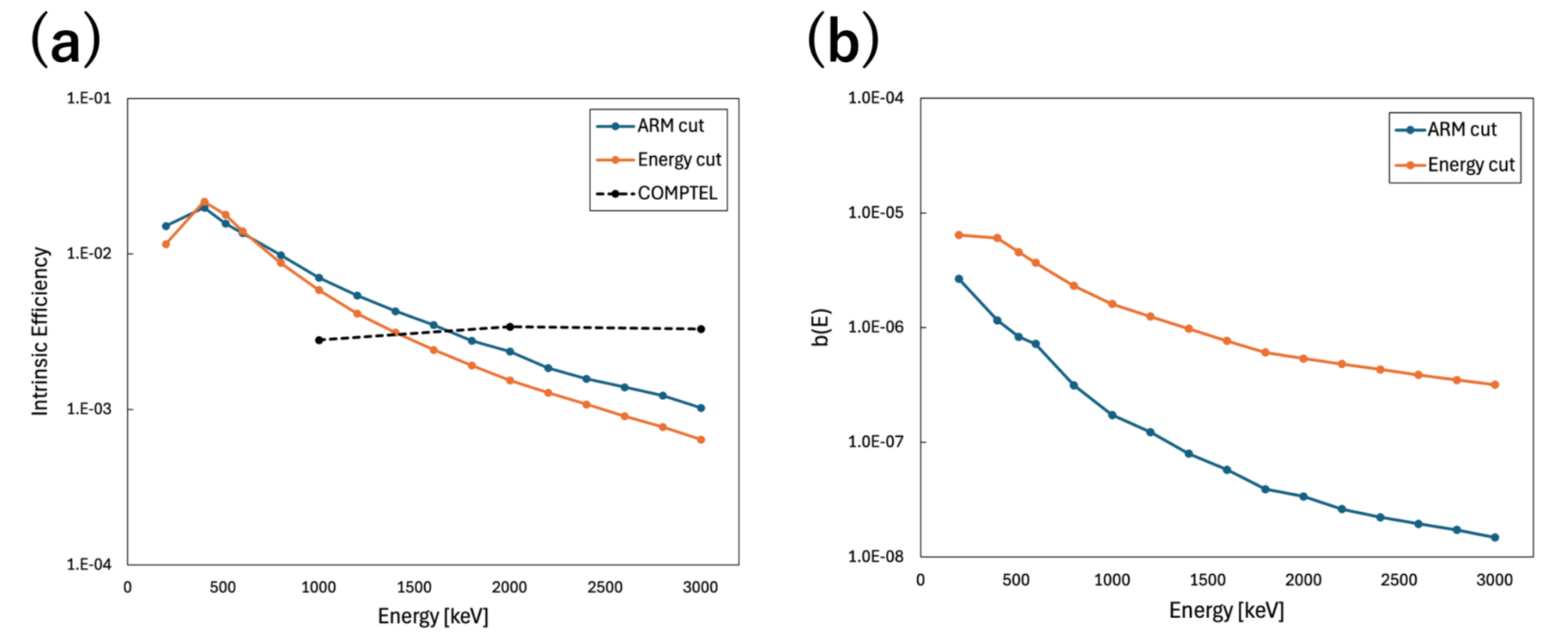}
\caption{Comparison of the (a) intrinsic efficiency $\eta(E)$  and 
(b) contaminated background flux  $b(E)$  for energy cut ($orange$) and ARM cut  ($cyan$) as a function of the incident 
gamma-ray energy.} 
    \label{fig:ItaComp}
\end{figure}

\begin{figure}[ht]
    \centering
    \includegraphics[width = 15.5cm]{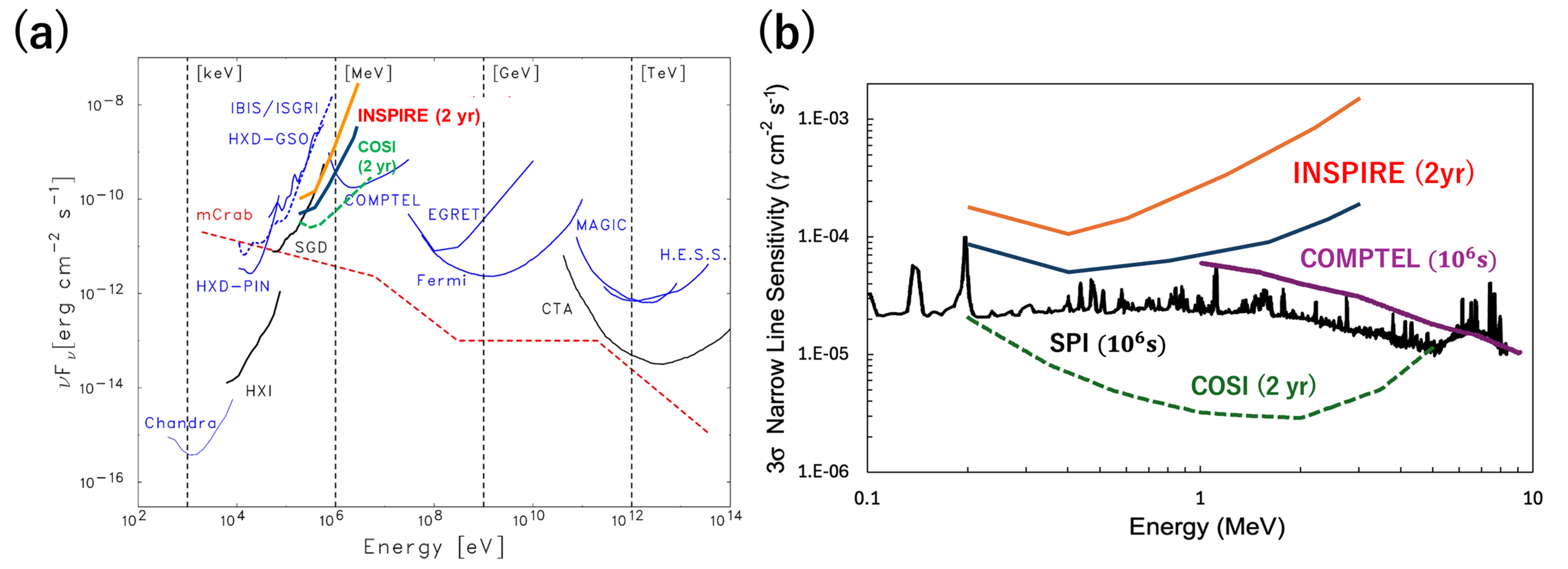}
\caption{Comparison of the (a) continuum sensitivity $S_{\rm C}(E)$, and 
(b) line sensitivity $S_{\rm L}(E)$ for energy cut ($orange$) and ARM cut ($cyan$) for $a$ = 1.} Note, sensitivity of INSPIRE over a 2-year observation corresponds to roughly 7$\times$10$^6$ sec along the Galactic plane.   The figure of the continuum sensitivity (a) was reconstructed from \cite{tak13} and \cite{tak14}. The lines representing the Chandra/ACIS-S, the Suzaku/HXD (PIN and GSO), INTEGRAL/IBIS (from the 2009 IBIS Observer’s Manual), and the Astro-H/HXI, and SGD are the 3$\sigma$ sensitivity curves for 100 ks exposures. The sensitivities of the COMPTEL and EGRET instruments correspond to the all-sky survey 
conducted over the entire lifetime of CGRO. The curve denoting Fermi-LAT reflect the pre-launch sensitivity, evaluated for a 5$\sigma$ detection limit at high Galactic latitudes, covering energy ranges of quarter-decade in a one-year dataset \cite{At09}. The curves for the MAGIC stereo system \cite{Ca11} and H.E.S.S. are given for a 5 $\sigma$ detection with more than 10 excess photons after a 50-hour exposure. The simulated CTA configuration  sensitivity curve for a 50-hour exposure at a
zenith angle of 20 degrees is taken from \cite{CTA11}. The figure of the line sensitivity (b) was reconstructed from \cite{to24}.
    \label{fig:sensitivity}
\end{figure}

In these equations, $\eta(E)$ for conventional energy cut 
is shown in Fig.\ref{fig:angeta}(b); the corresponding  $\eta(E)$ for ARM cut (pointing) is also added in Fig.\ref{fig:ItaComp}(a). 
In the case of the ARM cut with $a$ = 1, the peak of $\eta$ around 400~keV remains almost unchanged; however,  above 1~MeV, $\eta$ is 1.5 times larger than that of energy cut. This is because even a few  forward-scattered events in the BS-dominated region (see, Fig.\ref{fig:Ecut})  can be used in the ARM cut. In the case of 400~keV gamma rays, most of the events were forward-scattered; thus, they are less important.  However,  above 1~MeV, the contribution of such events would increase the 
sensitivity. In addition, Fig.\ref{fig:ItaComp}(b) compares 
$b(E)$ for the energy cut and ARM cut for $a$ = 1. $b(E)$ is  reduced approximately one order of magnitude 
in the case of ARM cut in the entire energy band. Because  both 
$S_{\rm C}(E)$ and $S_{\rm L}(E)$ contains $\eta(E)$ and $b(E)$ in 
the denominator and  numerator, the sensitivity is expected to  improve 
for pointing observations when ARM cut is applied.

Fig.\ref{fig:sensitivity} compares the sensitivities of the energy cut  and ARM cut for $a$ = 1.  In INSPIRE, a survey observation is planned in which the intersection line between the galactic plane and the satellite's orbit plane is always centered in the field of view during SSO observations. In this case, the observation time of the galactic plane will range from approximately 1.5 to 4 Msec/year depending on the galactic longitude, with about 3.5 Msec/year around the Galactic center. Therefore, the sensitivity for a 2-year observation, which is the minimum lifetime of GRAPHIUM, was calculated for a total exposure of 7 Msec.
The ARM cut significantly improves sensitivity, especially above 1~MeV, indicating that albedo gamma rays from outside the FOV may cause coincidences. However, such events were removed by checking the consistency between $\theta_E$ and $\theta_{\rm src}$. Notably, as shown in Fig.\ref{fig:sensitivity}(a) and (b), INSPIRE $-$ which has  dimensions and size approximately 1/100 of the COMPTEL $-$ achieves comparable sensitivity below 1~MeV for both continuum and line emission. Furthermore, when applying the ARM cut, the estimated line sensitivity closely approaches that of INTEGRAL/SPI  in the same energy range. The validity of this estimation will be discussed in $\S$4.2.

\begin{figure}[ht]
    \centering
    \includegraphics[width = 15.0cm]{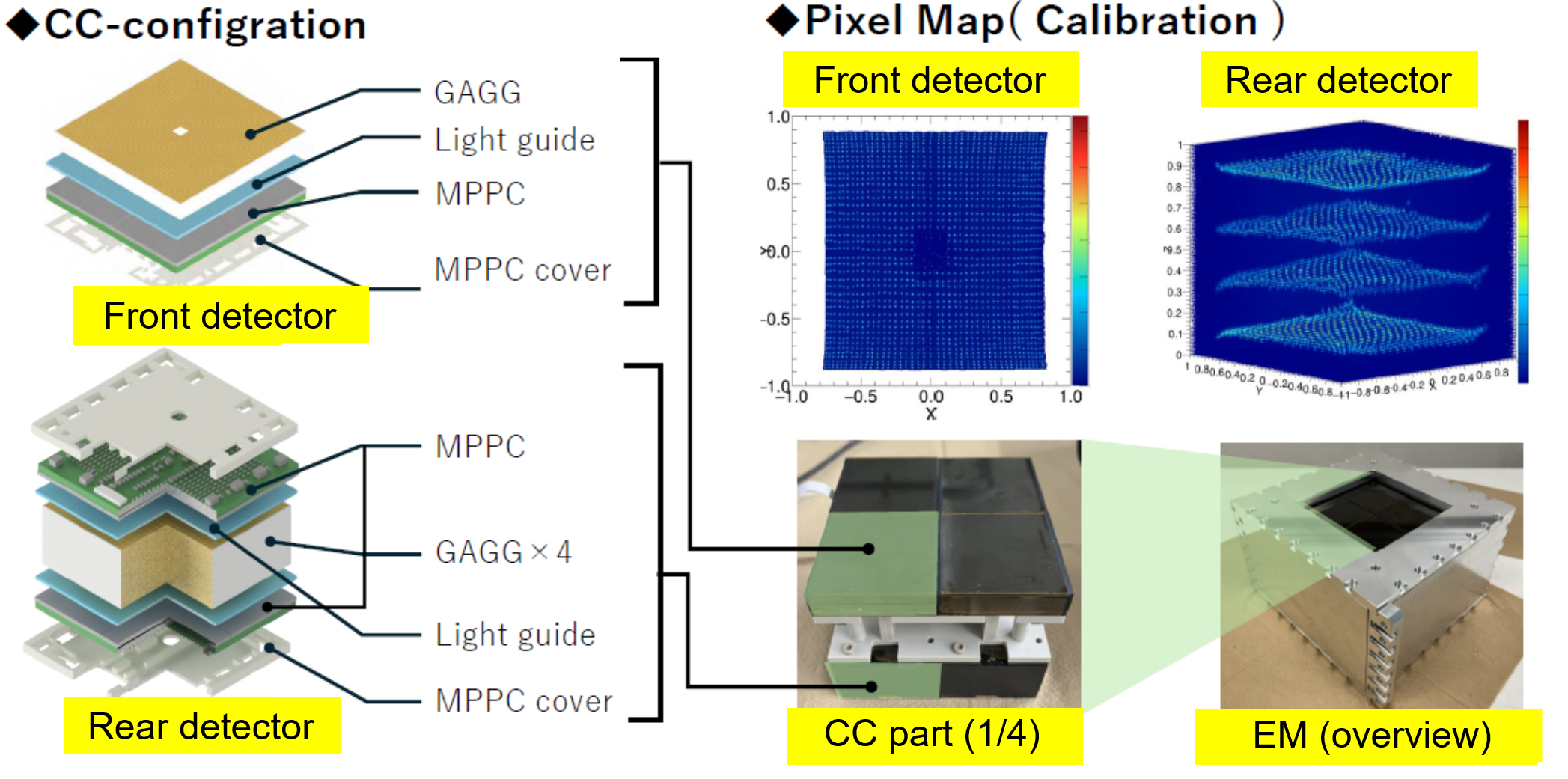}
\caption{Overview and internal structure of the EM of INSPIRE.} 
    \label{fig:EM}
\end{figure}

\subsection{Experiments using Engineering Model of INSPIRE}
\begin{figure}[ht]
    \centering
    \includegraphics[width = 10.0cm]{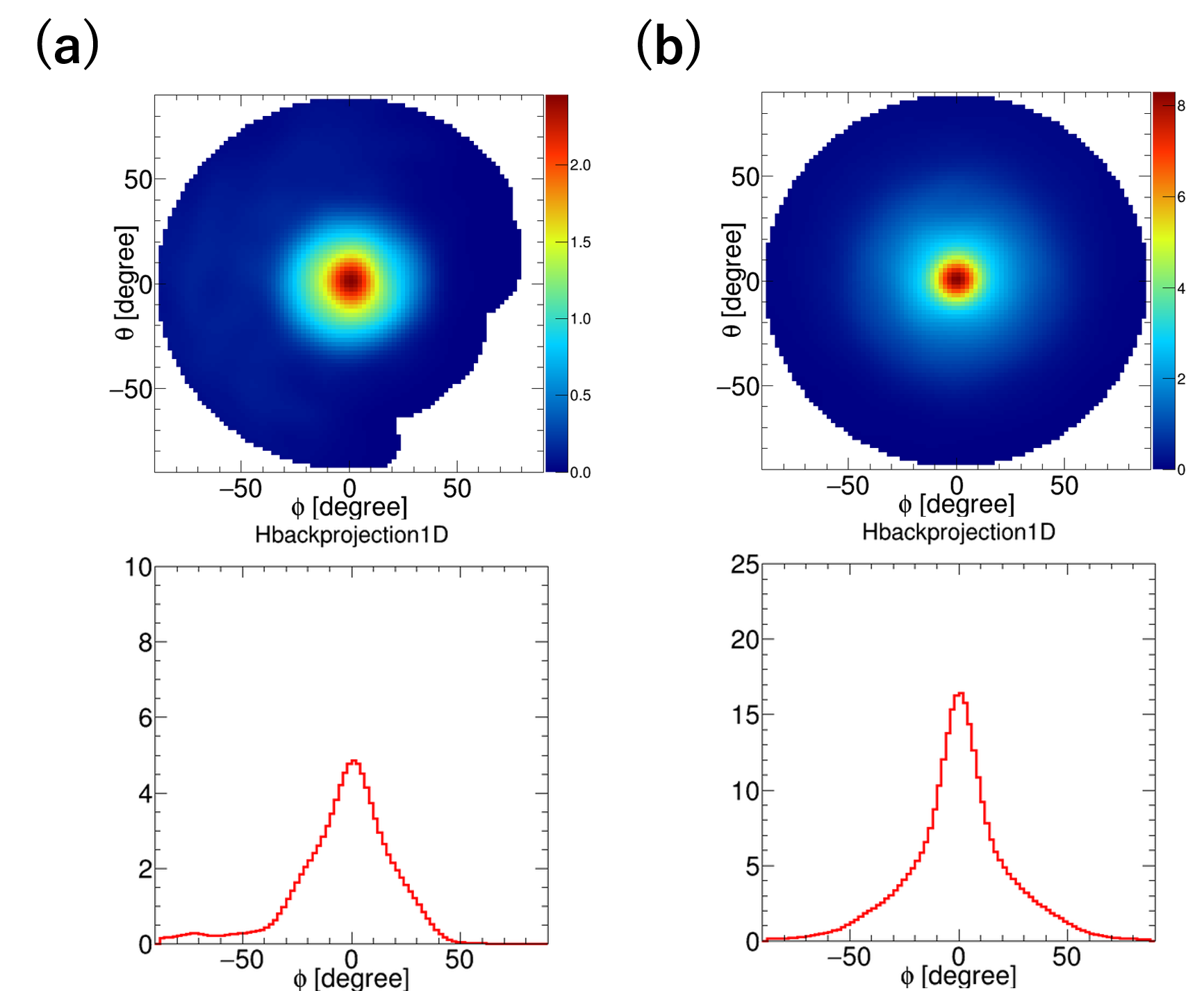}
\caption{Comparison of the Co-60 image (1333~keV) adopting (a) energy cut and 
(b) ARM cut ($a$=1; ARM = 3.75$^{\circ}$) and their 
one-dimensional projection.}     
\label{fig:Co-60}
\end{figure}

Finally, the effectiveness of the ARM cut proposed in this study was experimentally verified using the engineering model (EM; \cite{mor25}) of INSPIRE. The EM is a prototype in which only one of the four HCCs in INSPIRE is an actual Compton camera and the remaining three are replaced with dummy sensors with equivalent volumes and masses. The GAGG absorbers on the sides and BGO scintillators on the sides and bottom were implemented in the area adjacent to actual HCC in the EM. 
Fig.\ref{fig:EM} shows the appearance, detailed structure, and calibration data of the pixel map obtained using EM. Fig.\ref{fig:Co-60} shows a comparison of back-projection images obtained with energy cut and 
ARM cut, when 
Co-60 (1~MBq) is placed at the center of the FOV, 1.5~m away from the EM sensor. The measurement duration was 26 h. 
Applying the ARM cut ($a$ = 1) significantly improved the S/B ratio of the images, in terms of both peak intensity and FWHM,  primarily due to the suppression of escape events that do not satisfy ARM cut condition described in  Eqn.(2.3) .

\begin{figure}[ht]
    \centering
    \includegraphics[width = 14.0cm]{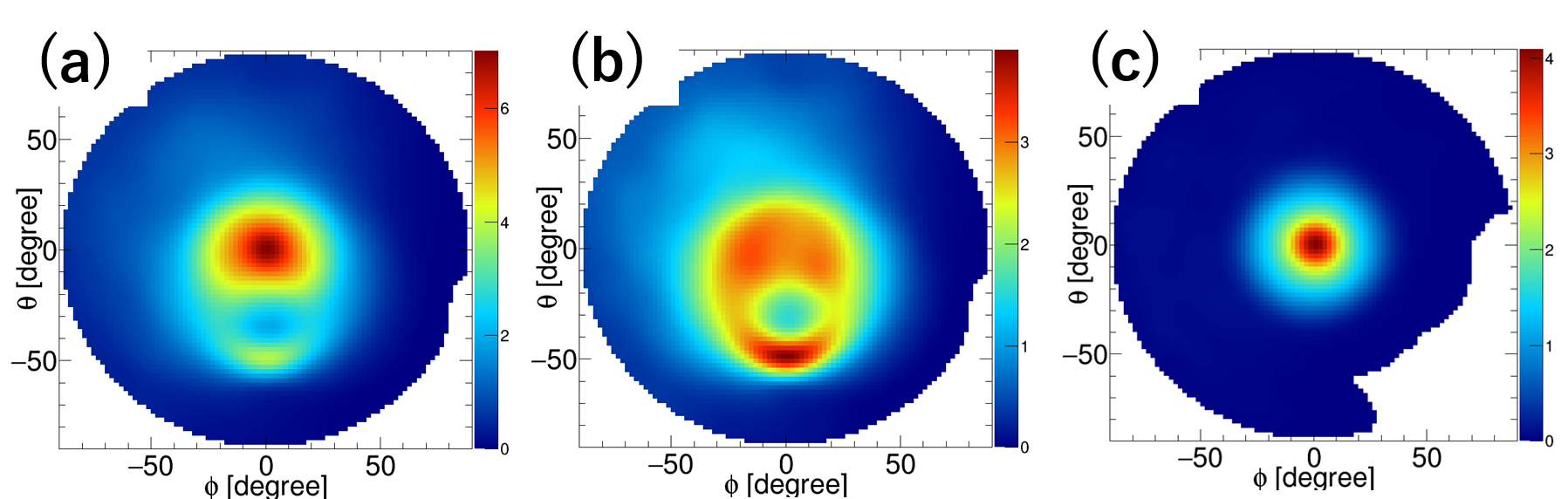}
\caption{Multiple source (Co-60; center, Cs-137; bottom) imaging with energy cut. (a) Image with energy cut OR'ed for both 662~keV (Cs-137) and 1333~keV (Co-60). 
(b) Image extracted with energy cut corresponds to Cs-137 only.
(c) Image extracted with energy cut corresponds to Co-60 only.} 
    \label{fig:multiE}
\end{figure}

To further demonstrate the effectiveness of the ARM cut, we 
captured images of multiple sources. Here, Co-60 (1~MBq) and 
Cs-137 (1~MBq) were placed at the center of FOV and  45$^{\circ}$ off the center, respectively. The images were 
acquired for 26 h. Fig.\ref{fig:multiE}(a) shows a back-projection image of a coincidence event with the energy cut corresponding to 
both Cs-137 (662~keV) and Co-60  (1333~keV) sources. Namely, for Cs-137, the energy 
ranges are 10 keV $<$$E_1$$<$ 160 keV and 612 keV $<$ $E_1$$+$$E_2$ $<$ 712~keV, whereas 
for Co-60, they are 10 keV $<$$E_1$$<$ 200 keV and 1290 keV $<$ $E_1$$+$$E_2$ $<$ 1390~keV, respectively.
In addition, images extracted by applying the energy cut  
either to the  Cs-137 or Co-60 sources are shown in Fig.\ref{fig:multiE}(b) and in Fig.\ref{fig:multiE}(c), respectively.
The positions of both were accurately identified by adopting the corresponding energy window. However, the image of 
Cs-137, Fig.\ref{fig:multiE}(b), was significantly degraded by escape events from Co-60 source, where only a part of the energy was deposited in the detector. Severe contamination prevented the extraction of the energy spectrum of each source from the corresponding image areas. This situation is more  challenging in actual astronomical observations, because two sources often exhibit similar power-law spectra, which make it virtually impossible 
to separate the source within the same FOV. 
In such a case,  simultaneous fitting of multiple sources are necessary.

\begin{figure}[ht]
    \centering
    \includegraphics[width = 14.0cm]{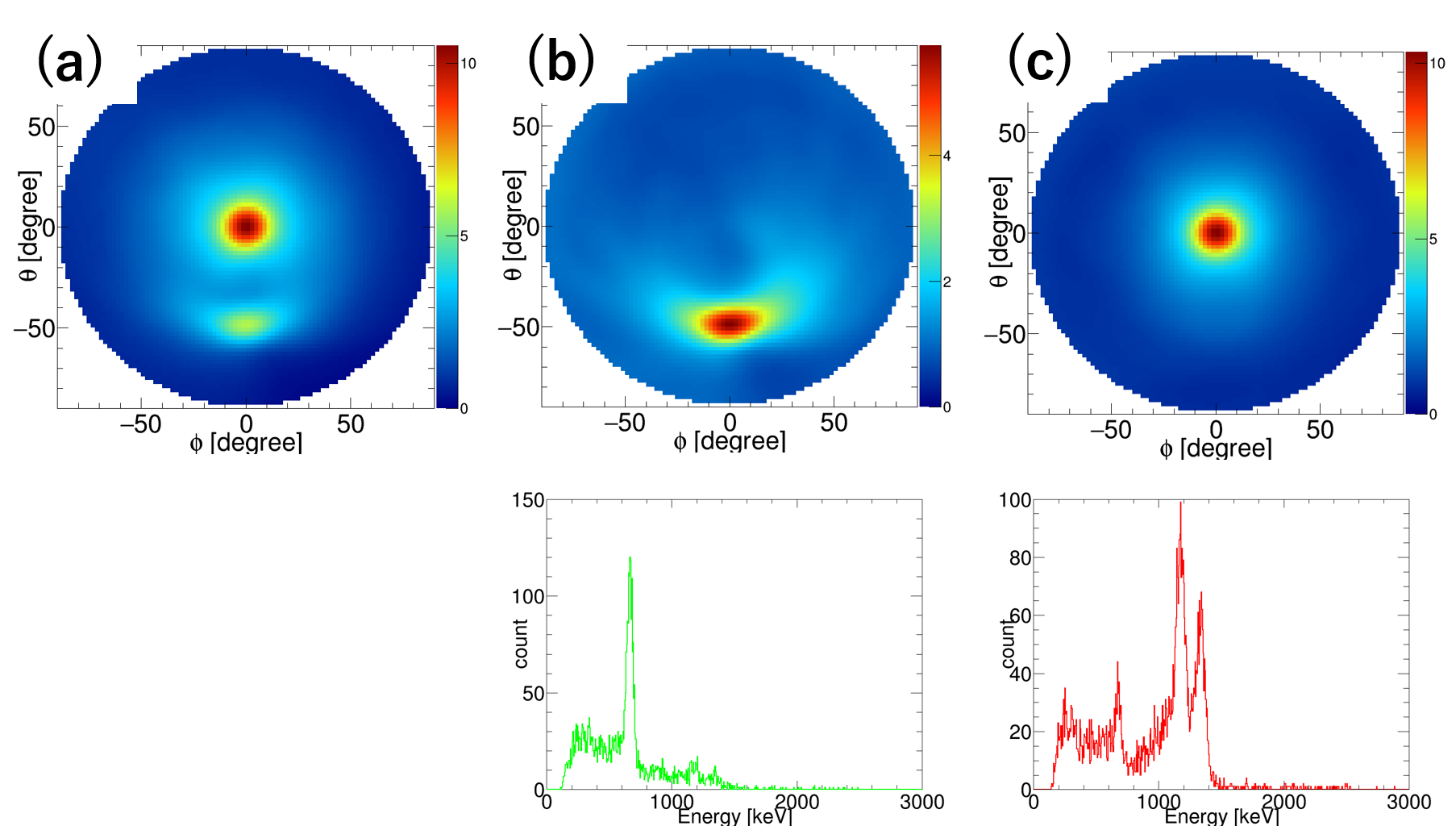}
\caption{Multiple source (Co-60; center, Cs-137; bottom) imaging with ARM cut and extraction of 
spectra from each sources.
(a) Image with ARM cut centered on both 662~keV (Cs-137) and 1333~keV (Co-60) positions. 
(b) Image  and spectrum extracted with ARM cut corresponds to Cs-137 position.
(c) Image and spectrum extracted with ARM cut corresponds to Co-60 position.}  
    \label{fig:multiA}
\end{figure}

In contrast, Fig.\ref{fig:multiA}(a) shows the back-projection image  corresponding to the ARM cut ($a$ = 1), where the two 
sources are clearly separated. Furthermore, the spectra corresponding to each source was  extracted by selecting the 
events that satisfied the ARM cut centered on each source, as shown in the bottom panel of Fig.\ref{fig:multiA}(a).
Fig.\ref{fig:multiA}(b) shows the image and spectra, assuming the positions of Cs-137. The image of Cs-137 
is unaffected by contamination with Co-60,  and the corresponding 662~keV photoelectric peak is clearly visible in the spectra.
Conversely, Fig.\ref{fig:multiA}(c) assumes that the positions of Co-60, which are unaffected by contamination from  neighboring Cs-137 source. Moreover, the peaks at 1173 and 1333 keV are clearly extracted from the spectra. Thus, performing 
the ARM cut allows the extraction of the spectra of multiple sources in the same FOV in the Compton camera image. 

\section{Discussion}
\subsection{Comparison with other approach using ARM cut}
In this paper, we revisited an ARM cut method $-$ a simple and practical method that can be easily applied  to
various  Compton cameras with different configurations.
Our simulations and experiments demonstrated that the method  effectively increased efficiency of the detector
and reduced the  background, thereby improving both the continuum and line sensitivity for 
pointing observations of astronomical sources.  Again, the idea of using ARM for event selection is not entirely new and has been discussed  in literature, but the method presented here emphasizes ease of implementation and broad applicability.  In the updated analysis of COMPTEL data, Kn\"odseler et al. (2022) developed a sophisticated software based on a dedicated plug-in for the GammaLib library, which is a community-developed toolbox for analyzing astronomical gamma-ray data.
The ARM cut is used to evaluate the residuals of the fit, which is performed using the maximum likelihood method in the Compton data space.
Therefore, the analysis implicitly assumed the ARM in a  convolutional way, such that events from a sky position 
($\alpha$, $\delta$) should be consistent with $\theta_E$ within a certain range;  however, energy dependence was not considered for the ARM distribution. In reality, the optimum range of ARM cut  should be varied as a function of energy to compile as much events as possible.

Although  the ARM cut proposed in this paper is easy to implement and may have broad applicability for background reduction,  statistical data analysis remains essential for imaging analysis, as shown in \cite{kn22}.  In this context, we reiterate again that 
while the ARM cut can be a powerful tool for  extracting spectra of known point sources, it 
cannot be use to determine the spacial distributions of unknown  gamma-ray sources.

Event selection using ARM (ARM cut) was also proposed for analyzing the Soft Gamma-ray Detector (SGD) onboard the Hitomi satellite \cite{ic16}. 
The SGD is a narrow field-of-view (FOV) Compton camera with the Si-CdTe detectors are embedded in a deep well of 
veto counter of the BGO scintillator. To achieve  
efficient Si/Cd detection with a narrow FOV of 5$^{\circ}$$\times$5$^{\circ}$, SGD observations were limited to 40$-$600 keV.  Moreover, fine collimators restrict the FOV to 0.6$^{\circ}$$\times$ 0.6$^{\circ}$.  The ARM cut was applied to reject events originating outside the FOV defined by BGO counter, but it was inefficient for gamma rays at low energies, as the ARM often exceeds the size of the FOV.  Although our proposed ARM cut appears similar,  it is applicable to any Compton cameras possessing a wide FOV and wider energy range. This is important because two major  advantages of a Compton camera are (1) its wide FOV and (2) the 
ability to observe gamma rays with energy greater than MeV.  For  pointing observations,  the flexible ARM cut can be applied in 
any energy range to effectively reduce the background,  regardless of the FOV of the detector. 

\subsection{About the estimation of sensitivity}
In Fig. \ref{fig:sensitivity}, we estimated the observational sensitivity for both the continuum and line components of INSPIRE. As described in $\S$ 3.2, the sensitivity calculations only considered contamination from the CXB and albedo gamma rays, without including effects caused by activation of the scintillator. Here, we briefly discuss why activation is not a significant concern for INSPIRE. Previous gamma-ray missions such as OSSE/CGRO and HXD/Suzaku used passive or active collimators to narrow the FOV but were non-imaging detectors. In those cases, contamination from CXB and albedo gamma rays was effectively suppressed; however, activation of the scintillator caused by ions (mainly protons) trapped in the SAA was a major source of background increase, degrading sensitivity. In contrast, INSPIRE has a much larger FOV of 3.1 sr, making the background contamination from within the FOV far greater than any internal background increase due to activation. Furthermore, the coincidence detection between the scatterer and absorber reduces such false events caused by activation. We also note that the GAGG scintillator exhibits relatively low activation compared to other scintillators, which further minimizes its impact on sensitivity \cite{yon18}.

 Next, we compare the line component sensitivity of INSPIRE with that of INTEGRAL/SPI \cite{di18}, which employs high-purity germanium (HPGe) detectors with excellent energy resolution. As shown in Eq. (3.2), the sensitivity is proportional to the square root of the detector’s energy resolution $\delta E$, the background flux $b(E)$, and inversely proportional to the square root of the detector area $A$. INTEGRAL/SPI has an area of approximately 500 cm$^2$, while INSPIRE’s area is 100 cm$^2$. Regarding energy resolution at around 1 MeV, INTEGRAL/SPI achieves $E/\delta E \simeq 450$, whereas INSPIRE’s $E/\delta E$ is approximately 20. Meanwhile, INTEGRAL’s orbit is an extremely elliptical trajectory, with a perigee at about 9,000 km and an apogee at 155,000 km. This causes it to pass through the Van Allen belts, resulting in background levels significantly higher than those in low Earth orbit (LEO). The background spectrum, compared to that of the Crab, is shown in Fig. 8 of Siegert et al. (2019). It indicates that the background around 1 MeV is more than 100 times higher than the albedo gamma-ray background. Furthermore, it is known that radiation damage to sensors at INTEGRAL's orbital altitude is one to two orders of magnitude higher than in LEO environments such as SSO \cite{t22}. Therefore, despite its smaller detector size and lower $E$/$\delta E$, INSPIRE is expected to achieve a line sensitivity close to that of INTEGRAL/SPI, thanks to its strategic orbit selection and design considerations.

In this context, we also compare INSPIRE's sensitivities with those of the COSI mission, currently under development by NASA\cite{to24}. COSI is planned to operate in LEO, with a detector area of approximately 256 cm², utilizing high-purity germanium (HPGe) detectors that achieve an energy resolution of about $E/\delta E$ $\simeq$ 450. In Fig.\ref{fig:sensitivity}, the expected sensitivity of COSI over a 2-year observation period is shown by the green dotted lines, alongside INSPIRE’s sensitivity for an all-sky survey during the same period, which corresponds to approximately 7$\times$10$^6$ seconds along the Galactic plane. The differences in sensitivity are primarily due to the detectors' energy resolution ($E/\delta E$). Additionally, INSPIRE’s scintillator has a thickness of approximately 20 mm (absorber), while COSI’s germanium detectors are 60 mm thick, leading to notable differences in detection capabilities at higher energies. Meanwhile, the GRAPHIUM satellite weighs only about one-sixth of COSI’s mass and has a development budget roughly one-hundredth of COSI’s. Though such ultra-small satellites face significant challenges due to limited resources such as weight, size, and power supply, they also have the potential to foster innovative ideas that can advance frontier science in the near future.

 \section{Conclusion}
We revisited a simple and effective method based on ARM cut that 
can be applied to Compton camera imaging.  Although the method is proposed 
for deep  observations of persistent point-like 
sources, like AGNs and pulsars, it is also suitable for transient events 
such as solar flares or GRBs,  whose positions are already known. 
In addition, we demonstrated that the ARM cut was useful for extracting multiple source spectra in the same FOV as the Compton camera. 
However, statistical data analysis remains essential for imaging studies where the positions and spatial extents of sources are unknown, such as in observations of large-scale diffuse emission associated with the Galactic plane. Currently, we are developing the flight model  of INSPIRE for the scheduled launch in 2027.




\acknowledgments
We thank an anonymous referee for his/her insightful comments to improve the manuscript.
This research was supported by JST ERATO (Grant Nos.JPMJER2102) 





\end{document}